\documentclass[runningheads,a4paper]{llncs}

\usepackage[english]{babel}
\usepackage[T1]{fontenc}
\usepackage{amssymb}

\makeatletter
\newif\ifpdfoutput
\@ifundefined{pdfoutput}%
  {\let\pdfoutput\@undefined}%
  {\ifcase\pdfoutput
     \let\pdfoutput\@undefined
   \else
     \pdfoutputtrue\fi
  }%
\makeatother

\ifpdfoutput
  \usepackage{lmodern}
\fi

\newcommand\vpspace{\ensuremath{\mathsf{VPSPACE}}}
\newcommand\vpspaceb{\ensuremath{\mathsf{VPSPACE}_{\mathsf{b}}}}
\newcommand\vpspacezero{\ensuremath{\mathsf{VPSPACE}^{\mathsf{0}}}}
\newcommand\vparzero{\ensuremath{\mathsf{VPAR}^0}}
\newcommand\pspace{\ensuremath{\mathsf{PSPACE}}}
\newcommand\vpzero{\ensuremath{\mathsf{VP}^{\mathsf{0}}}}
\newcommand\rr{\ensuremath{\mathbb{R}}}
\newcommand\cc{\ensuremath{\mathbb{C}}}
\newcommand\parr{\ensuremath{\mathsf{PAR}_{\rr}}}
\newcommand\parrzero{\ensuremath{\mathsf{PAR}_{\rr}^{\mathsf{0}}}}
\newcommand\pr{\ensuremath{\mathsf{P}_{\rr}}}
\newcommand\vp{\ensuremath{\mathsf{VP}}}
\newcommand\vpnb{\ensuremath{\mathsf{VP}_{\mathsf{nb}}}}
\newcommand\vpnbzero{\ensuremath{\mathsf{VP}^{\mathsf{0}}_{\mathsf{nb}}}}
\newcommand\vnp{\ensuremath{\mathsf{VNP}}}
\newcommand\vnpzero{\ensuremath{\mathsf{VNP}^{\mathsf{0}}}}
\newcommand\vnpnbzero{\ensuremath{\mathsf{VNP}^{\mathsf{0}}_{\mathsf{nb}}}}
\newcommand\poly{\ensuremath{\mathsf{poly}}}
\newcommand\zz{\ensuremath{\mathbb{Z}}}
\newcommand\ff{\ensuremath{\mathbb{F}}}
\newcommand\p{\ensuremath{\mathsf{P}}}
\newcommand\np{\ensuremath{\mathsf{NP}}}
\newcommand\nc{\ensuremath{\mathsf{NC}}}
\newcommand\unif{\ensuremath{\mathsf{Uniform\ }}}

\newcommand\parityhp{\ensuremath{\oplus\mathrm{HamiltonPath}}}
\newcommand\macaulay{\ensuremath{\mathrm{Mac}}}
\newcommand\mon{\ensuremath{\mathrm{Mon}}}
\newcommand\parityp{\ensuremath{\oplus\mathsf{P}}}
\newcommand\pu{\ensuremath{\mathsf{P}\mbox{-}\mathsf{uniform\ }}}

\begin{document}

\mainmatter
\title{VPSPACE and a Transfer Theorem\\ over the Reals}

\author{Pascal Koiran\and Sylvain Perifel}

\institute{LIP\thanks{UMR 5668 ENS Lyon, CNRS, UCBL, INRIA.}, 
\'Ecole Normale Sup\'erieure de Lyon.
{\tt [Pascal.Koiran,Sylvain.Perifel]@ens-lyon.fr}\\
\bigskip\today}


\maketitle

\begin{abstract}
We introduce a new class \vpspace\ of families of polynomials.
Roughly speaking, a family of polynomials is in \vpspace\ if its coefficients 
can be computed in polynomial space.
Our main theorem is that if (uniform, constant-free) 
\vpspace\ families can be evaluated efficiently
then the class \parr\ of decision problems that can be solved 
in parallel polynomial time over the real numbers collapses to \pr.
As a result, one must first be able to show that there are 
\vpspace\ families which are hard to evaluate 
in order to separate \pr\ from $\np_{\rr}$, or even from \parr.

\end{abstract}

\section{Introduction}

Two main categories of problems are studied in algebraic complexity theory:
evaluation problems and decision problems.
A typical example of an evaluation problem is the evaluation of the permanent
of a matrix, and it is well known that the permanent family is complete
for the class \vnp\ of ``easily definable'' 
polynomial families~\cite{valiant1979}.
Deciding whether a multivariate polynomial has a real root 
is a typical example of a decision problem. This problem is \np-complete
in the Blum-Shub-Smale model of computation 
over the real numbers~\cite{BCSS,BSS1989}.

The main purpose of this paper is to provide a transfer theorem connecting
the complexity of evaluation and decision problems.
This paper is therefore in the same spirit as~\cite{KoiPe06}.
In that paper, we showed
that if certain polynomials can be evaluated efficiently then
 certain decision problems become easy.
The polynomials considered in~\cite{KoiPe06} are those that can be written
as exponential-size products of polynomials that are easy to compute
(see~\cite{KoiPe06} for a precise definition) over some field $K$.
The decision problems under consideration are 
those that are in \np\ in the structure $(K,+,-,=)$,
in which multiplication is not allowed.

In the present paper we work with a larger class of polynomial families, which
we call \vpspace. Roughly speaking, a family of polynomials (of possibly
exponential degree) is in \vpspace\ if its coefficients can be evaluated in
polynomial space.  For instance, we show that resultants of systems of
multivariate polynomial equations form a \vpspace\ family.  Our main result is
that if (uniform, constant-free) \vpspace\ families can be evaluated
efficiently then the class \parr\ of decision problems that can be solved in
parallel polynomial time over the real numbers collapses to \pr.  This result
relies crucially on a combinatorial lemma due to
Grigoriev~\cite{grigoriev1999} and especially on its effective version,
recently established in~\cite{CJKPT2006}.  The class \parr\ plays roughly the
same role in the theory of computation over the reals as \pspace\ in discrete
complexity theory.  In particular, it contains $\np_{\rr}$~\cite{BCSS} (but
the proof of this inclusion is much more involved than in the discrete case).
It follows from our main result that in order to separate \pr\ from
$\np_{\rr}$, or even from \parr, one must first be able to show that there are
\vpspace\ families which are hard to evaluate.  This seems to be a very
challenging lower bound problem, but it is still presumably easier than
showing that the permanent is hard to evaluate.\\
{\em Organization of the paper}.  Section~\ref{sec_notions} recalls some
notions and notations from algebraic complexity (Valiant's model, the
Blum-Shub-Smale model).  The class \vpspace\ is defined in
Section~\ref{sec_vpspace} (both in a uniform and a nonuniform setting) and as
an example we show in Section~\ref{example} that resultants of multivariate
polynomial systems form a (uniform) \vpspace\ family. It is not necessary to
read Section~\ref{example} in order to understand the remainder of the
paper. Some closure properties of \vpspace\ are given in
Section~\ref{sec_closure}.

In Section~\ref{hypothesis}, the hypothesis that \vpspace\ families are easy
to evaluate is discussed. It is shown that (assuming the generalized Riemann
hypothesis) this hypothesis is equivalent to: $\vp = \vnp$ and $\p/\poly =
\pspace/\poly$.  The conjunction of these two equalities is an extremely
strong assumption: by results from~\cite{burgisser2000} (see~\cite{Koi04}), it
implies, assuming again GRH, that $\nc/\poly = \pspace/\poly$.  This
conjunction of equalities is still apparently consistent with our current
understanding of complexity theory.  We also discuss the uniform,
constant-free version of the hypothesis that \vpspace\ families are easy to
evaluate.  It turns out that this stronger hypothesis implies that $\pspace$
collapses to the polynomial-time uniform version of \nc.  Such a dramatic
collapse of complexity classes looks extremely unlikely, but as far as we know
it cannot be refuted with the current methods of complexity theory.

Finally, the last two sections of the paper are devoted
to the transfer theorem.  Section~\ref{sec_sign} deals with sign conditions,
an important tool from computational real algebraic geometry.  The transfer
theorem is stated at the beginning of Section~\ref{sec_input}, and proved
thereafter.

\section{Preliminaries}\label{sec_notions}

The notions of boolean complexity theory that we use are quite standard.
In the present section, we focus on algebraic complexity.

\subsection{The Blum-Shub-Smale Model} \label{BSS}

In contrast with boolean complexity, algebraic complexity deals with other
structures than $\{0,1\}$. In this paper we will focus on the ordered field
$(\rr,+,-,\times,\leq)$ of the real numbers. 
Although the original definitions of
Blum, Shub and Smale~\cite{BSS1989,BCSS} are in terms of 
uniform machines, 
we will follow~\cite{poizat1995} by using families of algebraic circuits to
recognize languages over $\rr$, that is, subsets of
$\rr^{\infty}=\bigcup_{n\geq 0}\rr^n$.

An algebraic circuit
is a directed
acyclic graph whose vertices, called gates, have indegree 0, 1 or 2.  An input
gate is a vertex of indegree 0.  An output gate is a gate of outdegree 0. We
assume that there is only one such gate in the circuit. Gates of indegree 2
are labelled by a symbol from the set $\{+,-,\times\}$. Gates of indegree 1,
called test gates, are labelled ``$\leq 0$?''. The size of a circuit $C$, in
symbols $|C|$, is the number of vertices of the graph.

A circuit with $n$ input gates computes a function from $\rr^n$ to $\rr$.  On
input $\bar u \in \rr^n$ the value returned by the circuit is by definition
equal to the value of its output gate.  The value of a gate is defined in the
usual way. Namely, the value of input gate number $i$ is equal to the $i$-th
input $u_i$.  The value of other gates is then defined recursively: it is the
sum of the values of its entries for a $+$-gate, their difference for a
$-$-gate, their product for a $\times$-gate.  The value taken by a test gate
is 0 if the value of its entry is $> 0$ and 1 otherwise.  We assume without
loss of generality that the output is a test gate. The value returned by the
circuit is therefore 0 or 1.

The class \pr\ is the set of languages $L \subseteq \rr^{\infty}$ such that
there exists a tuple $\bar a \in \rr^p$ (independent of $n$) and a \p-uniform
family of polynomial-size circuits $(C_n)$ satisfying the following condition:
$C_n$ has exactly $n+p$ inputs, and for any $\bar x \in \rr^n$, $\bar x \in L
\Leftrightarrow C_n(\bar x,\bar a)=1$.  The \p-uniformity condition means that
$C_n$ can be built in time polynomial in $n$ by an ordinary (discrete) Turing
machine.  Note that $\bar a$ plays the role of the machine constants
of~\cite{BCSS,BSS1989}.

As in~\cite{ChaKoi99}, we define the class \parr\ 
as the set of languages over \rr\ recognized by a 
\pspace-uniform
family of algebraic circuits of polynomial depth (and possibly exponential
size), with constants $\bar a$ as for \pr. 
Note at last that we could also define similar classes without constants
$\bar a$. We will use the superscript 0 to denote these constant-free classes,
for instance $\pr^0$ and $\parrzero$.

\subsection{Valiant's Model}

In Valiant's model, one computes polynomials instead of recognizing languages.
We thus use arithmetic circuits instead of algebraic circuits. A book-length
treatment of this topic can be found in~\cite{burgisser2000}.

An arithmetic circuit is the same as an algebraic circuit but test gates are
not allowed. That is to say we have indeterminates $x_1,\dots,x_{u(n)}$ as
input together with arbitrary constants of \rr; there are $+$, $-$ and
$\times$-gates, and we therefore compute multivariate polynomials.

The polynomial computed by an arithmetic circuit is defined in the usual way
by the polynomial computed by its output gate. Thus a family $(C_n)$ of
arithmetic circuits computes a family $(f_n)$ of polynomials,
$f_n\in\rr[x_1,\dots,x_{u(n)}]$. The class \vpnb\ defined in \cite{malod2003}
is the set of families $(f_n)$ of polynomials computed by a family $(C_n)$ of
polynomial-size arithmetic circuits, i.e., $C_n$ computes $f_n$ and there
exists a polynomial $p(n)$ such that $|C_n|\leq p(n)$ for all $n$. We will
assume without loss of generality that the number $u(n)$ of variables is
bounded by a polynomial function of $n$. The subscript ``$\mathsf{nb}$''
indicates that there is no bound on the degree of the polynomial, in contrast
with the original class \vp\ of Valiant where a polynomial bound on the degree
of the polynomial computed by the circuit is required. Note that these
definitions are nonuniform. The class \unif\vpnb\ is obtained by adding a
condition of polynomial-time uniformity on the circuit family, as in
Section~\ref{BSS}.

The class \vnp\ is the set of families of polynomials defined by an
exponential sum of \vp\ families. More precisely, $(f_n(\bar x))\in\vnp$ if
there exists $(g_n(\bar x,\bar y))\in\vp$ and a polynomial $p$ such that
$|\bar y| = p(n)$ and
$f_n(\bar x) = \sum_{\bar\epsilon\in\{0,1\}^{p(n)}}g_n(\bar x,
\bar\epsilon).$

We can also forbid constants from our arithmetic circuits in unbounded-degree
classes, and define constant-free classes. The only constant allowed is 1 (in
order to allow the computation of constant polynomials). As for classes of
decision problems, we will use the superscript 0 to indicate the absence of
constant: for instance, we will write \vpnbzero\ (for bounded-degree classes,
we are to be more careful; see~\cite{malod2003}).

Note at last that arithmetic circuits are at least as powerful as boolean
circuits in the sense that one can simulate the latter by the former. Indeed,
we can for instance replace $\neg u$ by $1-u$, $u\wedge v$ by $uv$, and $u\vee
v$ by $u+v-uv$. This proves the following classical lemma.

\begin{lemma}\label{lem_simulation}
  Any boolean circuit $C$ can be simulated by an arithmetic one of size at
  most $3|C|$, in the sense that on boolean inputs, both circuits output the
  same value.
\end{lemma}

\section{The Class VPSPACE}\label{sec_vpspace}

\subsection{Definition}

We fix an arbitrary field $K$. The definition of \vpspace\ will be stated in
terms of \emph{coefficient function}. A monomial $x_1^{\alpha_1}\cdots
x_{u(n)}^{\alpha_{u(n)}}$ is encoded in binary by $\alpha =
(\alpha_1,\dots,\alpha_{u(n)})$ and will be written $\bar x^{\alpha}$.

\begin{definition}
  Let $(f_n)$ be a family of multivariate polynomials with integer
  coefficients. The coefficient function of $(f_n)$ is the function $a$ whose
  value on input $(n,\alpha,i)$ is the $i$-th bit $a(n,\alpha,i)$ of the
  coefficient of the monomial $\bar x^\alpha$ in $f_n$. Furthermore,
  $a(n,\alpha,0)$ is the sign of the coefficient of the monomial $\bar
  x^{\alpha}$. Thus $f_n$ can be written as
  $$f_n(\bar x) = \sum_{\alpha}\Bigl((-1)^{a(n,\alpha,0)}\sum_{i\geq 1}
  a(n,\alpha,i)2^{i-1}\bar x^{\alpha}\Bigr).$$
\end{definition}

The coefficient function is a function $a:\{0,1\}^*\rightarrow\{0,1\}$ and can
therefore be viewed as a language. This allows us to speak of the complexity
of the coefficient function.

\begin{definition}
  The class \unif\vpspacezero\ is the set of all families $(f_n)$ of
  multivariate polynomials $f_n\in K[x_1,\dots,x_{u(n)}]$ satisfying the
  following requirements:
  \begin{enumerate}
  \item the number $u(n)$ of variables is polynomially bounded;
  \item the polynomials $f_n$ have integer coefficients;
  \item the size of the coefficients of $f_n$ is bounded by $2^{p(n)}$ for
    some polynomial $p$;
  \item the degree of $f_n$ is bounded by $2^{p(n)}$ for some polynomial $p$;
  \item the coefficient function of $(f_n)$ is in \pspace.
  \end{enumerate}
\end{definition}

We have chosen to define first \unif\vpspacezero, a uniform class without
constants, because this is the main object of study in this paper.  In keeping
with the tradition set by Valiant, however, the class \vpspace, defined in
Section~\ref{nonuniform_class}, is nonuniform and allows for
arbitrary constants.

\subsection{An Alternative Characterization}

Let \unif\vparzero\ be the class of families of polynomials computed by a 
\pspace-uniform family of constant-free arithmetic circuits 
of polynomial depth (and possibly
exponential size). This in fact characterizes \unif\vpspacezero.

\begin{proposition}\label{prop_alternate}
The two classes $\unif\vpspacezero$ and $\unif\vparzero$ are equal.
\end{proposition}

\begin{proof}
Let $(f_n)$ be a $\unif\vpspacezero$ family.
In order to compute $f_n$ by an arithmetic circuit of polynomial depth,
we compute all its monomials in parallel and sum them 
in a divide-and-conquer-fashion.
The resulting family of arithmetic circuits is uniform due to the uniformity 
condition on $(f_n)$.

  For the converse, take an arithmetic circuit of
  polynomial depth. We show that we can build a boolean circuit of polynomial
  depth which takes as input the encoding $\alpha$ of a monomial and computes
  the coefficient of $\bar x^{\alpha}$. We proceed by induction, computing the
  coefficient of $\bar x^{\alpha}$ for each gate of the original arithmetic
  circuit. For the input gates, this is easy. For a $+$-gate, it is enough to
  add both coefficients. For a gate $a\times b$, we compute in parallel the
  sum of the $cd$ over all the monomials $\bar x^\beta$ and $\bar x^\gamma$
  such that $\beta+\gamma=\alpha$, where $c$ is the coefficient of $\bar
  x^\gamma$ in the gate $a$, and $d$ the coefficient of $\bar x^\beta$ in the
  gate $b$. The whole boolean circuit remains uniform and of polynomial
  depth. Therefore, the coefficient function is in \pspace\ by the ``parallel
  computation thesis''.
\qed
\end{proof}

We see here the similarity with \parr, which by definition are those languages
recognized by uniform algebraic circuits of polynomial depth. But of course
there is no test gate in the arithmetic circuits of \unif\vpspacezero.

\subsection{An Example}\label{example}

Algebraic geometry is a natural source of examples for the study of
polynomials from a computational point of view. For instance, 
the Hilbert polynomial is studied in~\cite{BL2005} 
from the point of view of discrete complexity theory.
Here we study a different example:
the computation of the resultant of a system of multivariate
polynomials. 
A system of $n+1$ homogenous equations in $n+1$ complex variables 
has a nontrivial solution if and only if its resultant 
is zero. 
We sketch the construction of the resultant below. 
More details can be found for instance in \cite{macaulay1916} 
or \cite{canny1988}.

Let $f_1,\dots,f_{n+1}\in\cc[X_0,\dots,X_n]$ be a system of $n+1$ homogeneous
polynomials. 
The resultant consists in the quotient of the determinants of two 
matrices $M$ and $M'$:
\begin{equation} \label{quotient}
R=\frac{\det M}{\det M'}
\end{equation}
where the coefficients of $M$ are among those of the $f_i$'s, and $M'$ is a
submatrix of $M$. The matrix $M$ is called Macaulay's matrix (a generalization
of Sylvester's for two univariate polynomials) and is described 
as follows.
Let $d_i$ be the degree of $f_i$ and
$d=1+\sum_{i=1}^{n+1}(d_i-1)$. Denote by $\mon_d$ the set of all monomials in
$X_0,\dots,X_n$ of degree $d$: the cardinal of $\mon_d$ is $N={d+n\choose d}$.

The matrix $M$ has $N$ rows and $N$ columns, both indexed by the elements of
$\mon_d$. The row corresponding to the monomial $\bar x^\alpha$ represents the
polynomial
$$\frac{\bar x^\alpha}{x_i^{d_i}}f_i\mbox{, where }i=\min\{j;x_j^{d_j} 
\mbox{ divides } \bar x^\alpha\}.$$ 
Finally, the submatrix $M'$ consists in the rows and columns of
$M$ that are ``not reduced'', see~\cite{canny1988}.
What we will compute is not
the resultant $R$ itself but rather a multiple of it, namely $\det
M$. Whenever $\det M'\neq 0$, this does not change anything if we are only
concerned by the vanishing of $R$.

From now on, we will assume for simplicity that all the $d_i$ are equal.
We will let $n$ go to infinity, but the common value $\delta$ 
of the $d_i$ will remain constant.
A system $(f_1,\dots,f_{n+1})$ of $n+1$ homogeneous polynomials 
of degree $\delta$ in
$n+1$ variables is encoded by the list of the coefficients
of the polynomials, i.e.,~by $k(n+1)$ variables $(a_{1,1},\dots,
a_{1,k},a_{2,1}, \dots,a_{n+1,k})$ where $k={n+\delta \choose \delta}$ 
is the number of monomials of degree $\delta$ in $n+1$ variables. 
Note that $k$ is polynomial in $n$ for any fixed $\delta$.

The matrix
$\macaulay_n^{\delta}(f_1,\dots,f_{n+1})$
is then defined as the Macaulay matrix
$M$ of $(f_1,\dots, f_{n+1})$. This matrix is of size $n+d \choose d$,
where $d=1+(n+1)(\delta-1)$. 
This is exponential in $n$ as soon as $\delta \geq 2$.
Computing the determinant of $M$ can be
done by a circuit of depth polylogarithmic in the size of $M$, thus polynomial
in $n$. 
The above considerations then prove
the following proposition.
\begin{proposition}
 For any fixed $\delta$,  the family $(\det(\macaulay_n^{\delta}))$ 
(the determinant of the Macaulay matrix
  of a system of $n+1$ homogeneous polynomials of degree $\delta$ 
in $n+1$ variables) is in
  \unif\vpspacezero.
\end{proposition}
Likewise, the determinants of the matrices $M'$ in~(\ref{quotient}) form a
\unif \vpspacezero\ family.

\subsection{Closure Properties}\label{sec_closure}

The following lemma is clear
from Proposition~\ref{prop_alternate}.
\begin{lemma}\label{lem_sum_prod}
\unif \vpspacezero\ is closed under big sums and big products.
\end{lemma}

We can even make sums and products over a set more complicated than $\{0,1\}$,
as proven in the following lemma.
\begin{lemma}\label{lem_sum_prod_pspace}
  Let $A$ be a language in \pspace, $(f_n(\bar x,\bar y))$ a family in
  \unif\vpspacezero\ and $p(n)$ a polynomial, where $|\bar y|=p(n)$. Then the
  families $(g_n(\bar x))$ and $(h_n(\bar x))$ defined as follows are in
  \unif\vpspacezero.
  $$g_n(\bar x) = \sum_{\bar\epsilon\in A^{=p(n)}}f_n(\bar
  x,\bar\epsilon)\mbox{ and }
  h_n(\bar x) = \prod_{\bar\epsilon\in A^{=p(n)}}f_n(\bar x,\bar\epsilon).$$
\end{lemma}

\begin{proof}
  It is enough to use Lemma~\ref{lem_sum_prod} since we have
  $$\sum_{\bar\epsilon\in A^{=p(n)}}f_n(\bar
  x,\bar\epsilon) = \sum_{\bar\epsilon\in\{0,1\}^{p(n)}}
  \chi_A(\bar \epsilon)f_n(\bar x,\bar\epsilon)\mbox{, and}$$
  $$\prod_{\bar\epsilon\in A^{=p(n)}}f_n(\bar x,\bar\epsilon) =
  \prod_{\bar\epsilon\in\{0,1\}^{p(n)}}
  [\chi_A(\bar\epsilon)f_n(\bar x,\bar\epsilon)+(1-\chi_A(\bar\epsilon))],$$
  where $\chi_A$, the characteristic function of $A$, is in \unif\vpspacezero\
  by Lemma~\ref{lem_simulation} and Proposition~\ref{prop_alternate} since $A$
  is decided by a \unif family of boolean circuits of polynomial depth.\qed
\end{proof}

\subsection{The Nonuniform Class VPSPACE} \label{nonuniform_class}

Let us now define the nonuniform classes \vpspacezero\ and \vpspace. Note that
the only difference between \vpspacezero\ and \unif\vpspacezero\ is the
nonuniformity of the coefficient function.
\begin{definition} \label{vpspace}
  The class \vpspacezero\ is the set of all families $(f_n)$ of
  multivariate polynomials $f_n\in K[x_1,\dots,x_{u(n)}]$ satisfying the
  following requirements:
  \begin{enumerate}
  \item the number $u(n)$ of variables is polynomially bounded;
  \item the polynomials $f_n$ have integer coefficients;
  \item the size of the coefficients of $f_n$ is bounded by $2^{p(n)}$ for
    some polynomial $p$;
  \item the degree of $f_n$ is bounded by $2^{p(n)}$ for some polynomial $p$;
  \item the coefficient function of $(f_n)$ is in \pspace/\poly.
  \end{enumerate}
  Now, the class \vpspace\ is the set of all families $(f_n(\bar x))$ of
 multivariate polynomials $f_n\in K[x_1,\dots,x_{u(n)}]$ such that there
 exist a family $(g_n(\bar x,\bar y))\in\vpspacezero$ together with a family
  of tuples of constants $(\bar a^{(n)})$ satisfying for all $n$:
 $$f_n(\bar x)=g_n(\bar x,\bar a^{(n)}).$$
\end{definition}

We introduce temporarily a degree-bounded version of \vpspace: this will prove
useful for comparing \vpspace\ to \vp\ and \vnp\ since the degree of the
polynomials in these last two classes are polynomially bounded. A family
$(f_n)$ of polynomials is in $\vpspaceb^0$ if $(f_n)\in\vpspace^0$ 
and the size of
the coefficients as well as the degree of $f_n$ are polynomially
bounded. 
The class \vpspaceb\ is then defined from $\vpspaceb^0$ in the same way as
\vpspace\ is defined from $\vpspace^0$ in Definition~\ref{vpspace}.
This new class is
interesting for our purpose due to the following two lemmas.
\begin{lemma}\label{lem_equiv}
  $$\vpspaceb=\vp\iff\vpspace=\vpnb.$$
\end{lemma}
\begin{proof}
Assume first that $\vpspace=\vpnb$, and 
take a family $(f_n)\in\vpspaceb$. Since
  $\vpspaceb\subset\vpspace$, $(f_n)$ is in fact in \vpnb\ by hypothesis. Now,
  since the degree of $(f_n)$ is polynomially bounded, $(f_n)\in\vp$.

  For the converse,
take a family $(f_n)\in\vpspace$: remember
  that it can be written as $f_n(\bar x)=g_n(\bar x,\bar a^{(n)})$ for some
  constants $\bar a^{(n)}$ and $(g_n(\bar x,\bar y))\in\vpspacezero$.
  For convenience, let us rename the $u(n)$ variables of $g_n$ by
  $v_1,\dots,v_{u(n)}$, thus we have:
  $$g_n(\bar v) = \sum_{\alpha}\Bigl((-1)^{a(n,\alpha,0)}\sum_{i=1}^{2^{p(n)}}
  a(n,\alpha,i)2^{i-1}\bar v^{\alpha}\Bigr),$$
  where $a$ is in \pspace/\poly. In this expression, $p(n)$ is a polynomial
  and $2^{p(n)}$ bounds the size of the coefficients  as well as the degree of
  $g_n$. In order to use the hypothesis, we have to somehow define a family
  $(h_n)\in\vpspaceb^0$ that will ``simulate'' $(g_n)$. Let us define
  $$\Bigl(h_n(z_{1,1},\dots,z_{1,p(n)},
  z_{2,1},\dots,z_{u(n),p(n)}, w_1,\dots,w_{p(n)})\Bigr),$$
  where intuitively the variable $z_{i,j}$ is to replace $v_i^{2^j}$ in $g_n$,
  and $w_i$ will take the value $2^{2^i}$. More formally, $h_n$ is defined as
  follows:
  \begin{itemize}
  \item replace $v_i^k$ in $g_n$ by $\prod_{j\in J_k}z_{i,j}$, where the set
    $J_k$ consists of the bits set to 1 in the binary representation of $k$;
  \item replace the coefficient $2^{i-1}$ in the term $\sum_{i=1}^{2^{p(n)}}
    a(n,\alpha,i)2^{i-1}$ of $g_n$ by $\prod_{j\in J_{i-1}} w_j$, where the
    set $J_{i-1}$ consists of the bits set to 1 in the binary representation
    of $i-1$.
  \end{itemize}
  The degree of $h_n$ is then polynomially bounded and all the coefficients
  are among $-1$, 0 and 1. Note furthermore that the coefficient function is
  still in \pspace. Therefore $(h_n)\in\vpspaceb^0$, thus $(h_n)\in\vp$ 
by hypothesis. It remains to replace $z_{i,j}$ by $v_i^{2^j}$ and $w_i$ by
  $2^{2^i}$ to show that $(g_n(\bar v) =g_n(\bar x,\bar y))\in\vpnb$, and then
  to replace $\bar y$ by the original constants in order to show that
  $(f_n)\in\vpnb$.\qed
\end{proof}

\begin{lemma}\label{lem_vpspaceb_vnp}
  \vpspaceb\ contains \vnp.
\end{lemma}

\begin{proof}
  Let $(HC_n)$ be the family defined by
  $$HC_n(x_{1,1},\dots,x_{1,n},x_{2,1},\dots,x_{n,n})=
  \sum_{\sigma}\prod_{i=1}^n x_{i,\sigma(i)}$$
  where the sum is taken over all $n$-cycles $\sigma$ over
  $\{1,\dots,n\}$. This polynomial counts the number of Hamilton cycles in a
  graph given by its adjacency matrix. $(HC_n)$ is \vnp-complete,
  see~\cite{valiant1979} or~\cite{malod2003}. Since \vpspaceb\ is closed under
  $p$-projections and contains $HC_n$, the lemma follows.\qed
\end{proof}

\section{On the Hypothesis that VPSPACE has Small Circuits} \label{hypothesis}

In this section,
we investigate some consequences of the hypotheses $\vpspace=\vpnb$ 
and $\unif\vpspacezero = \unif\vpnbzero$.
\begin{proposition}\label{prop_equivalence}
Under the generalized Riemann hypothesis (GRH),
$$\vpnb=\vpspace\iff[\p/\poly=\pspace/\poly\mbox{ and }\vp=\vnp].$$
Moreover, the implication
from right to left holds even without GRH.
\end{proposition}
\begin{proof}
Assume first that $\p/\poly=\pspace/\poly$ and $\vp=\vnp$.
By Lemma~\ref{lem_equiv},
  the equality $\vpspace=\vpnb$ is equivalent to the degree-bounded analogue
  $\vpspaceb=\vp$. Let $(f_n)\in\vpspaceb$: its coefficient function is in
  \pspace/\poly, thus in \p/\poly\ by our assumption. Since the set of
  coefficient functions of \vnp\ families contains \parityp/\poly\ (see
  \cite{burgisser2000}), hence \p/\poly, $(f_n)$ is in fact in \vnp. By our
  assumption again, it is in \vp.

For the converse, assume now that
  $\vpspace=\vpnb$.  Again, this is equivalent to $\vpspaceb=\vp$. 
Hence $\vnp=\vp$ since $\vp\subseteq\vnp\subseteq\vpspaceb$ by
  Lemma~\ref{lem_vpspaceb_vnp}. It remains to show that a language $A$ in
  \pspace/\poly\ belongs in fact to \p/\poly. $A$ is recognized by a
  \p/\poly-uniform family of polynomial-depth boolean circuits, and by
  Lemma~\ref{lem_simulation} and Proposition~\ref{prop_alternate} there exists
  a family $(f_n)\in\vpspace$ such that on any 
boolean input $\bar x\in\{0,1\}^n$,
  $f_n(\bar x)\in\{0,1\}$ and $f_n(\bar x)=1$ if and only if $\bar x\in A$.

  By our assumption, $(f_n)\in\vpnb$, thus there exists a family of
  polynomial-size arithmetic circuits $(C_n)$, with arbitrary constants, that
  computes $(f_n)$. In order to evaluate these circuits on boolean inputs with
  boolean circuits, the problem now is to eliminate the constants. We
  proceed as in \cite{burgisser2000}. Let $\bar y$ be the constants for the
  circuit $C_n$, and call $g_n(\bar X,\bar Y)$ the polynomial computed by
  $C_n$ where the constants are replaced by the new variables $\bar Y$. Thus
  $g_n(\bar X, \bar y)=f_n(\bar X)$, therefore the system $S$ of equations in
  $\bar Y$ defined by
  $$S=\left(g_n(\bar x,\bar Y)=f_n(\bar x)\right)_{\bar x\in\{0,1\}^n}$$
  has a solution $\bar y$ over \cc. All the equations in this system have
  integer coefficients, degree bounded by $2^{q(n)}$ and weight by
  $2^{2^{q(n)}}$ for some polynomial $q$, where the weight of a polynomial is
  the sum of the absolute value of its coefficients.

  By Theorem~4.4 of~\cite[p.~64]{burgisser2000}, assuming GRH there exists a
  prime number $p\leq 2^{n^2q(n)}$ such that $S$ has a solution over
  $\ff_p$. There indeed exists such a $p\leq a$ as soon as
  $$\frac{\pi(a)}{d^{O(n)}}>\sqrt{a}\log(wa),$$
  where $d$ and $w$ are bounds on the degree and weight of the equations
  respectively, and $\pi(a)$ is the number of primes $\leq a$. Thus there
  exists a polynomial-size arithmetic circuit over $\ff_p$ computing the
  polynomial $g_n(\bar X,\bar y')$ and this polynomial takes the same values
  as $f_n(\bar X)$ on boolean inputs.

  Note that the size of $p$ is polynomial, and a solution $\bar y'$ of this
  system $S$ over $\ff_p$ also has polynomial size. Therefore a
  polynomial-size boolean circuit working modulo $p$ can now easily compute
  the value of $g_n(\bar X, \bar y')$ over $\ff_p$. This boolean circuit has
  the same value on boolean inputs as $f_n$. Hence $A\in\p/\poly$, and the
  announced result is proved.\qed
\end{proof}

We now turn in the next proposition to the most uniform version of the
hypothesis, which is stronger than that of
Proposition~\ref{prop_equivalence}. For the proof, we need two definitions
from~\cite{MRK1988} and~\cite{malod2003}. 
\begin{definition}
  The \emph{formal degree} of an arithmetic circuit $C$ is the formal degree
  of its output gate, where the formal degree of a gate is defined
  recursively:
  \begin{itemize}
  \item the formal degree of an input gate is 1;
  \item the formal degree of a $+$-gate or a $-$-gate is the maximum of the
    formal degrees of its inputs;
  \item the formal degree of a $\times$-gate is the sum of the formal degrees
    of its inputs.
  \end{itemize}
\end{definition}
\begin{definition}\label{def_vpzero}
  The class \vpzero\ is the set of families of polynomials computed by a
  family of constant-free (i.e.~using only 1 as a constant) polynomial-size
  arithmetic circuits of polynomial formal degree.
\end{definition}

The following proposition is similar to
Proposition~\ref{prop_equivalence}, but in a uniform setting and without
assuming the generalized Riemann hypothesis. It is not clear whether the
assumption $\unif\vpnbzero=\unif\vnpnbzero$ in this proposition can be
replaced by the assumption $\unif\vpzero=\unif\vnpzero$.
\begin{proposition}\label{prop_equivalence2}
$\unif\vpnbzero=\unif\vpspacezero$ if and only if
$$\p=\pspace\mbox{ and }\unif\vpnbzero=\unif\vnpnbzero.$$
\end{proposition}
\begin{proof}
  Assume first that $\p=\pspace$ and $\unif\vpnbzero=\unif\vnpnbzero$. Take
  a family $(f_n)\in\unif\vpspacezero$. Its coefficient function is in
  \pspace, hence in \p\ by assumption. The sum of the monomials with their
  coefficients is therefore in $\unif\vnpnbzero$. Thus
  $(f_n)\in\unif\vpnbzero$ by assumption.

  For the converse, let us first show that $\p=\pspace$. Let $A$ be a \pspace\
  language: it is decided by a uniform family of polynomial-depth boolean
  circuits. By Lemma~\ref{lem_simulation} and
  Proposition~\ref{prop_alternate}, we obtain a family of polynomials
  $(f_n)\in\unif\vpspacezero$ that agrees with the boolean circuits on boolean
  inputs, i.e.,
  $$\forall \bar x\in\{0,1\}^n,f_n(\bar x)\in\{0,1\}\mbox{ and }
  [f_n(\bar x)=1\Longleftrightarrow\bar x\in A].$$

  By our assumption, $(f_n)\in\unif\vpnbzero$ so that there exists a uniform
  family of polynomial-size arithmetic circuits that computes $(f_n)$. Of
  course, on boolean inputs such circuits can be evaluated in polynomial time
  (working modulo 2 to avoid overflows). This implies that $\pspace=\p$.

  Now, the proof of $\unif\vnpnbzero=\unif\vpnbzero$ is clear since
  $\unif\vpnbzero\subseteq\unif\vnpnbzero\subseteq\unif\vpspacezero$.\qed
\end{proof}

We can now prove a consequence of the hypothesis
$\unif\vpspacezero = \unif\vpnbzero$.
\begin{proposition}
  $$\unif\vpspacezero = 
\unif\vpnbzero\Longrightarrow\pspace=\pu\nc.$$
\end{proposition}
\begin{proof}
  By Proposition~\ref{prop_equivalence2}, the hypothesis already implies
  $\p=\pspace$. Let us now prove that $\parityp\subseteq\pu\nc$ under the
  hypothesis that $\unif\vpspacezero = \unif\vpnbzero$. It is enough to show
  that the
  \parityp-complete language \parityhp\ (the problem of deciding whether there
  is an odd number of Hamilton paths in a graph,
  see~\cite[p.~448]{papadimitriou1994}) is in \pu\nc. For a graph given by its
  boolean adjacency matrix $(a_{i,j})$ (where $a_{i,j}=1$ iff there is an edge
  between $i$ and $j$), the number of Hamilton paths is
  $$\sum_{1\leq j<k\leq n}\ \sum_{\sigma\in S_{j,k}}\ \prod_{i=1}^{n-1}
  a_{i,\sigma(i)},$$
  where $S_{j,k}$ is the set of all the $n$-cycles $\sigma\in\mathcal{S}_n$
  beginning in $j$ and ending in $k$ ($j$ is different from $k$ in order to
  count paths in the graph and not cycles, and $j$ is smaller than $k$ in order
  not to count twice each path, which would trivialize the
  problem \parityhp). The polynomial
  $$p_n(x_{1,1},\dots,x_{1,n},x_{2,1},\dots,x_{n,n})=\sum_{j<k}\ \sum_{\sigma\in
    S_{j,k}}\prod_{i=1}^{n-1} x_{i,\sigma(i)}$$ therefore outputs the number
  of Hamilton paths on the boolean encoding $x_{1,1}\dots x_{1,n}x_{2,1}\dots
  x_{n,n}$ of a graph $G$. This family of polynomials $(p_n)$ is easily seen
  to be in \unif\vpspacezero, has polynomially bounded degree, and its
  evaluation modulo 2 provides the answer to the question
  ``$G\in\parityhp?$''.

  By our assumption, $(p_n)\in\unif\vpnbzero$ so that there exists a \p-uniform
  family of polynomial-size arithmetic circuits $(C_n)$ that computes
  $(p_n)$. We are going to build a family of circuits $(D_n)$ that computes a
  family of polynomials $(q_n)\in\vpzero$ such that on boolean inputs, $p_n$
  and $q_n$ have the same parity. Note that despite the polynomial bound on
  its degree, $(p_n)$ needs not be already in \vpzero\ because the formal
  degree of $C_n$ needs not be polynomial (indeed, constants of exponential
  size might be computed by $C_n$). This is why we cannot directly evaluate
  $C_n$ in parallel with the algorithm of~\cite{MRK1988}.

  The idea here is that we can compute only the remainder modulo 2 of the
  constants because we are only interested in the result modulo 2. $D_n$ is
  then built from $C_n$ as follows. First, note that $p_n$ has degree
  $n-1$. We compute each homogeneous component separately: each gate $\alpha$
  of $C_n$ is split into $n-1$ gates $\alpha_1,\dots,\alpha_{n-1}$, the gate
  $\alpha_i$ computing the homogeneous component of degree $i$ of
  $\alpha$. The homogeneous components of degree 0 (i.e.~the constants) are
  not computed, only their remainder modulo 2 is taken into account. In other
  words, we replace an even constant by the constant 0, and an odd one by
  1. The \p-uniformity remains because we can compute in polynomial time the
  the constants modulo 2. The last step of $D_n$ is to compute the sum of
  the homogeneous components of the output gate.

  It is easy and well known how to compute these homogeneous components at
  each step, while keeping a polynomial circuit size: we merely discard the
  homogeneous components of degree $>n-1$. With this construction, it is clear
  that $p_n$ and $q_n$ coincide modulo 2, that the construction is \p-uniform,
  and that the formal degree of $D_n$ is at most $n-1$ because there is no
  constant in the circuit any more. Hence $(q_n)\in\vpzero$.

  In order to decide \parityhp, we therefore only have to compute the value of
  $q_n$ modulo 2 on the given input, that is, to evaluate a \p-uniform circuit
  of polynomial size $s(n)$ and polynomially bounded formal degree
  $n-1$. Theorem 5.3 of \cite{MRK1988} tells us that such a circuit can be
  evaluated modulo 2 by a logspace-uniform algorithm in parallel time
  $O(\log(s(n))\log(ns(n)))$, i.e.~$O(\log(n)^2)$, and with $O(n^2)$
  processors, thus placing \parityp\ in \pu$\nc^{\mathsf{2}}$.

  Hence, assuming that $\unif\vpspacezero = \unif\vpnbzero$ we have proved
  that
  $$\pspace=\p\subseteq\parityp\subseteq\pu\nc^{\mathsf{2}}.$$

  Note that this construction does not seem to be logspace uniform because
  evaluating the constants modulo 2 is a \p-complete problem.

  Since we construct a circuit family which is only polynomial-time uniform,
  one could also use the construction of~\cite{VSBR83} instead of the parallel
  algorithm of \cite{MRK1988}. Indeed, as pointed out in~\cite{MRK1988}, the
  construction of~\cite{VSBR83} can be performed in polynomial time.  \qed
\end{proof}

\begin{remark}
  Despite its unlikeliness, the separation ``$\pspace\neq\pu\nc$'' is not
  known to hold to the authors' knowledge 
(by contrast, $\pspace$ can be separated from logspace-uniform \nc\ thanks to
the space hierarchy theorem).
\end{remark}

\section{Sign Conditions}\label{sec_sign}

\subsection{Definition}

Given are $s$ polynomials $f_1,\dots,f_s\in\zz[x_1,\dots,x_n]$. A sign
condition is merely an $s$-tuple $S\in\{-1,0,1\}^s$. Intuitively, the $i$-th
coordinate of $S$ represents the sign of $f_i$: $-1$ for $<0$, 0 for 0, and 1
for $>0$. Accordingly, the sign condition of a point $\bar x\in\rr^n$ is the
tuple $S\in\{-1,0,1\}^s$ such that $S_i=-1$ if $f_i(\bar x)<0$, $S_i=0$ if
$f_i(\bar x)=0$ and $S_i=1$ if $f_i(\bar x)>0$.

Of course some sign conditions are not realizable, in the sense that the
polynomials can nowhere take the corresponding signs (think for instance of
$x^2+1$ which can only take positive values over \rr). We say that a sign
condition is \emph{satisfiable} if it is the sign condition of some $\bar
x\in\rr^n$ and we call $N$ the number of satisfiable sign conditions. The key
result detailed in the next section is that among all possible sign
conditions, there are few satisfiable ones (i.e.~$N$ is small), and there
exists a polynomial space algorithm to enumerate them all.

\subsection{A PSPACE Algorithm for Sign Conditions}

The following theorem will prove to be a central tool in our proofs. The bound
on the number of satisfiable sign conditions follows from the Thom-Milnor
bounds~\cite{milnor1964} (see Grigoriev~\cite[Lemma~1]{grigoriev1988}); the
enumeration algorithm is from Renegar~\cite[Prop.~4.1]{renegar1992part1}.

\begin{theorem}\label{th_renegar}
  Let $f_1,\dots,f_s\in\zz[x_1,\dots,x_n]$ be $s$ polynomials of maximal
  degree $d$, and whose coefficients have bit size $\leq L$. Then:
  \begin{enumerate}
  \item there are $N=(sd)^{O(n)}$ satisfiable sign conditions;
  \item there is an algorithm using work space $(\log L)[n\log(sd)]^{O(1)}$
    which, on input $(f_1,\dots,f_s)$ in dense representation, and $(i,j)$ in
    binary, outputs the $j$-th component of the $i$-th satisfiable sign
    condition.
  \end{enumerate}
\end{theorem}
If $S$ is the $i$-th satisfiable sign condition produced by this enumeration
algorithm, we say that the \emph{rank} of $S$ is $i$ (the rank is therefore
merely the index of the sign condition in the enumeration). Note that if
$d=2^{n^{O(1)}}$, $s=2^{n^{O(1)}}$ and $L=2^{n^{O(1)}}$ as will be the case,
then the work space of the algorithm is polynomial in $n$.

\subsection{Enumerating all Possibly Tested Polynomials}\label{sec_enum}

In the execution of an algebraic circuit, the values of some polynomials at
the input $\bar x$ are tested to zero. If two points $\bar x$ and $\bar y$
have the same sign condition with respect to all polynomials possibly tested
to zero, then they will either both belong to the language, or both be outside
of it: indeed the results of all the tests will be the same during the
execution of the circuit. Therefore we can handle sign conditions
(i.e.~boolean words) instead of algebraic inputs.

Note that in order to find the sign condition of the input $\bar x$, we have
to be able to enumerate in polynomial space all the polynomials that can ever
be tested to zero in some computation of an algebraic circuit. This is done as
in~\cite[Th.~3]{CK1997}.

\begin{proposition}\label{prop_slice}
  Let $C$ be a constant-free algebraic circuit with $n$ variables and of depth
  $d$.
  \begin{enumerate}
  \item The number of different polynomials possibly tested to zero in some
    computation of $C$ is $2^{d^2O(n)}$.
  \item There exists an algorithm using work space $(nd)^{O(1)}$ which, on
    input $C$ and integers $(i,j)$ in binary, outputs the $j$-th bit of the
    $i$-th of these polynomials.
  \end{enumerate}
\end{proposition}

\begin{proof}
    $C$ is sliced in levels corresponding to the depth of the gates: input gates
  are on the level 0 and the output gate is the only one on level $d$.

  Suppose that the results of the tests of the levels 0 to $i-1$ are fixed: we
  can then compute all the polynomials tested at level $i$. Since our agebraic
  circuits have fan-in at most 2, there are at most $2^{d-i}$ gates on level
  $i$ of $C$: in particular, at most $2^{d-i}$ polynomials can be tested
  on level $i$. But the degree of a polynomial computed at level $i$ is at
  most $2^i$ and the size of its coefficients is
  $(nd)^{O(1)}2^i$. Therefore, by Theorem~\ref{th_renegar} there are at most
  $(2^d)^{O(n)}$ possible outcomes for the tests of level $i$, and they are
  moreover enumerable in space $(nd)^{O(1)}$. Therefore we can compute all the
  $(2^d)^{O(n)}$ possible outcomes of all the tests of level $i$ and proceed
  inductively. This gives an algorithm using work space $(nd)^{O(1)}$
  for enumerating all the polynomials that can possibly be tested in an
  execution of the circuit. Since there are $2^{dO(n)}$ possible outcomes at
  each level, the total number of polynomials for the whole circuit (that is,
  for $d$ levels) is $(2^{dO(n)})^d=2^{d^2O(n)}$, as claimed in the statement
  of the proposition.  \qed
\end{proof}

Note that this proposition can also be useful when our algebraic circuit is
not constant-free: it is enough to replace the constants by fresh
variables. The only risk is indeed to take more polynomials into account since
we have replaced specific constants by generic variables.

\section{A Transfer Theorem}\label{sec_input}

In this section we prove our main result.
\begin{theorem}\label{th_transfer}
$\unif \vpspacezero = \unif\vpnbzero \Longrightarrow \parrzero = \pr^0.$
\end{theorem}
Note that the collapse of the constant-free class
\parrzero\ to
$\pr^0$ implies the collapse of
\parr\ to
\pr:
just replace constants by new variables in order to transform a \parr\
problem into a \parrzero\ problem, and then replace these variables 
by their orignal values in order to transform a $\pr^0$ problem 
into a \pr\ problem.

Let $A\in\parrzero$: it is decided by a uniform family $(C_n)$ of 
constant-free 
algebraic circuits of polynomial depth. 
For convenience, we fix $n$ and work with
$C_n$.
For the proof of Theorem~\ref{th_transfer} we will need to find 
the sign condition of the input $\bar
x$ with respect to the polynomials $f_1,\dots,f_s$ of
Proposition~\ref{prop_slice}, that is to say, with respect to all the
polynomials that can be tested to zero in an execution of $C_n$. We denote 
by $N$
the number of satisfiable sign conditions with respect to $f_1,\dots,f_s$.

Note that most of the forthcoming results depend on the polynomials
$f_1,\dots,f_s$, therefore on the choice of $C_n$. For instance, once $C_n$
and $f_1,\dots,f_s$ are chosen, the satisfiable sign conditions are fixed and
we will speak of the $i$-th satisfiable sign condition without referring
explicitly to the polynomials $f_1,\dots,f_s$.

In order to find the sign condition of the input, we will give a
polynomial-time algorithm which tests some \vpspace\ family for zero. Here is
the formalized notion of a polynomial-time algorithm with \vpspace\ tests.
\begin{definition}\label{def_algo_tests}
  A polynomial-time algorithm with \unif \vpspacezero\ tests is a \unif
  \vpspacezero\ family $(f_n(x_1,\dots,x_{u(n)}))$ together with a uniform
  constant-free family $(C_n)$ of polynomial-size algebraic circuits endowed
  with special test gates of indegree $u(n)$, whose value is $1$ on input
  $(a_1,\dots,a_{u(n)})$ if $f_n(a_1,\dots,a_{u(n)})\leq 0$ and $0$ otherwise.
\end{definition}
Observe that a constant number of \unif\vpspacezero\ families can be used in the
preceding definition instead of only one: it is enough to combine them all in
one by using ``selection variables''. The following Theorem~\ref{th_number2}
is the main result en route to showing the transfer theorem. It is proved via
successive lemmas in Sections~\ref{sec_trunc} to~\ref{sec_recovering}: we
proceed as in~\cite{grigoriev1999} but constructively.
\begin{theorem}\label{th_number2}
  There is a polynomial-time algorithm with \unif\vpspacezero\ tests that, on
  input $\bar x$, computes the rank of the sign condition of $\bar x$ with
  respect to $f_1,\dots,f_s$.
\end{theorem} 

\subsection{Truncated Sign Conditions}\label{sec_trunc}

A truncated sign condition is merely an element $T$ of $\{0,1\}^s$. Contrary
to full sign conditions, only the two cases $=0$ and $\neq 0$ are
distinguished. We define in a natural way the truncated sign condition $T$ of
a point $\bar x$: $T_i=0$ if and only if $f_i(\bar x)=0$.

Of course, there are fewer satisfiable truncated sign conditions than full
ones, and of course there exists a polynomial space algorithm to enumerate
them. Furthermore, truncated sign conditions can be viewed as subsets of
$\{1,\dots,s\}$ (via the convention $k\in T\iff T_k=1$), therefore enabling us
to speak of inclusion of truncated sign conditions.

We fix an order $\leq_T$ compatible with inclusion and easily computable in
parallel, e.g.~the lexicographic order. Let us call $T^{(i)}$ the $i$-th
satisfiable truncated sign condition with respect to this order.

\begin{lemma}\label{lem_trunc1}
  There is an algorithm using work space polynomial in $n$ which, on input
  $(f_1,\dots,f_s)$ in dense representation, and $(i,j)$ in binary, outputs
  the $j$-th component of $T^{(i)}$ (the $i$-th satisfiable truncated sign
  condition with respect to $\leq_T$).
\end{lemma}

\begin{proof}
  It is enough to use the algorithm of Theorem~\ref{th_renegar}, followed by a
  fast parallel-sorting procedure, for instance Cole's parallel merge-sort
  algorithm~\cite{cole1988}.\qed
\end{proof}

Note that the truncated sign condition of the input $\bar x$ is the maximal
truncated satisfiable sign condition $T$ satisfying $\forall
i,T_i=1\Rightarrow f_i(\bar x) \neq 0$. 
Hence we have to find a maximum. This will
be done by binary search.

\begin{lemma}\label{lem_trunc2}
  There is a \unif\vpspacezero\ family $(g_n)$ of polynomials satisfying, for
  real $\bar x$ and boolean $i$,
  $$g_n(\bar x,i) =
  \prod_{j\leq i}\Bigl(\sum_{k\not\in T^{(j)}}f_k(\bar x)^2\Bigr).$$
\end{lemma}

\begin{proof}
  Lemma~\ref{lem_trunc1} asserts that deciding whether $k\not\in T^{(j)}$ is
  in \pspace. Then we use twice Lemma~\ref{lem_sum_prod_pspace} (once for the
  sum and once for the product).\qed
\end{proof}

\begin{proposition}\label{prop_trunc}
  There is a polynomial-time algorithm with \unif\vpspacezero\ tests which on
  input $\bar x$ outputs the rank $m$ of its truncated sign condition
  $T^{(m)}$.
\end{proposition}

\begin{proof}
The algorithm merely consists in performing a binary search thanks to the
  polynomials of Lemma~\ref{lem_trunc2}: if the truncated sign condition of
  the input $\bar x$ is $T^{(m)}$, then $\prod_{j\leq i}\bigl(\sum_{k\not\in
    T^{(j)}}f_k(\bar x)^2\bigr)=0$ if and only if $m\leq i$. By making $i$
  vary, we find $m$ in a number of steps logarithmic in the number of
  satisfiable truncated sign conditions, i.e.~in polynomial time.\qed
\end{proof}

\subsection{Binary Search for the Full Sign Condition}

We say that a (full) sign condition $S$ is compatible with the truncated sign
condition $T$ if $\forall i, T_i=0\Leftrightarrow S_i=0$ (i.e.~they agree for
``$=0$'' and for ``$\neq 0$''). Let $N'$ denote the number of (full)
satisfiable sign conditions compatible with the truncated sign condition of
the input $\bar x$. Obviously, $N'\leq N$. The following lemma is
straightforward after Lemma~\ref{lem_trunc1} and Theorem~\ref{th_renegar}.

\begin{lemma}\label{lem_full1}
  There is an algorithm using work space polynomial in $n$ which, on input
  $(i,j,k)$, ouputs the $j$-th bit of the $i$-th satisfiable sign condition
  compatible with $T^{(k)}$.
\end{lemma}

Since we know the truncated sign condition of $\bar x$ after running the
algorithm of Proposition~\ref{prop_trunc}, we know which polynomials vanish at
$\bar x$. We can therefore discard the zeros in the (full) compatible
satisfiable sign conditions. Hence we are now concerned with two-valued sign
conditions, that is, elements of $\{-1,1\}^{s'}$ with $s'\leq s$. In what
follows arithmetic over the field of two elements will be used, hence it will
be simpler to consider that our sign conditions have values among $\{0,1\}$
instead of $\{-1,1\}$: 0 for $>0$ and 1 for $<0$. Thus sign conditions are
viewed as vectors over $\{0,1\}$, or alternately as subsets of
$\{1,\dots,s'\}$. The set $\{0,1\}^{s'}$ is endowed with the inner product
$u.v=\sum_i u_iv_i (\mbox{mod } 2)$, and we say that $u$ and $v$ are
orthogonal whenever $u.v=0$ (see~\cite{CJKPT2006}).

The following proposition from~\cite{CJKPT2006} will be useful. It consists in
an improvement of the result of~\cite{grigoriev1999}: first
(and most importantly),
 it is constructive, and second, 
the range $[N'/2-\sqrt{N'}/2,N'/2+\sqrt{N'}/2]$ here
is much better than the original one $[N'/3,2N'/3]$.
\begin{proposition}\label{prop_half}
  Let $V$ be a set of $N'$ vectors of $\{0,1\}^{s'}$.
  \begin{enumerate}
  \item There exists a vector $u$ orthogonal to at least $N'/2-\sqrt{N'}/2$
    and at most $N'/2+\sqrt{N'}/2$ vectors of $V$.
  \item Such a vector $u$ can be found on input $V$ by a logarithmic space
    algorithm.
  \end{enumerate}
\end{proposition}

Our aim is to find the sign condition of $\bar x$. We will use
Proposition~\ref{prop_half} in order to divide the cardinality of the search
space by two at each step. This is based on the following observation: if
$u\in\{0,1\}^{s'}$, the value of the product $\prod_{j\in u}f_j(\bar x)$ is
negative if the inner product of $u$ and the sign condition of $\bar x$ is 1,
and is positive otherwise. The idea is then to choose $u$ judiciously so that
the number of satisfiable sign conditions having the same inner product with
$u$ as the sign condition of $\bar x$ is halved at each step. Therefore, in a
logarithmic number of steps, the sign condition of $\bar x$ will be uniquely
determined. This gives the following algorithm for finding the sign condition
of $\bar x$.
\begin{itemize}
\item Let $E$ be the set of all the satisfiable sign conditions.
\item While $E$ contains more than one element, do
  \begin{itemize}
  \item Find by Proposition~\ref{prop_half} a vector $u$ orthogonal to at
    least $|E|/2-\sqrt{|E|}/2$ and at most $|E|/2+\sqrt{|E|}/2$ vectors of
    $E$.
  \item Let $b$ be the result of the test ``$\prod_{j\in u}f_j(\bar x)<0$?''.
  \item Let the new $E$ be the set of all sign conditions in $E$ which have
    inner product $b$ with $u$.
  \end{itemize}
\item Enumerate all the satisfiable sign conditions and find the one that
  produces exactly the same results as in the loop: this is the sign condition
  of $\bar x$.
\end{itemize}
Note that the number of steps is $O(\log N')$, which is polynomial in
$n$. The last step of this algorithm (namely, recovering the rank of the sign
condition of $\bar x$ from the list of results of the loop) is detailed in
Section~\ref{sec_recovering}.

We now show how to perform this algorithm in polynomial time with \unif
\vpspacezero\ tests. The main technical difficulty is that according to
Definition~\ref{def_algo_tests} we can use only one \vpspace\ family, whereas
we want to make adaptive tests. We therefore have to store the intermediate
results of the preceding tests in some variables $\bar c$ (a ``list of
choices'') of the \vpspace\ polynomial. Proposition~\ref{prop_half} shows
that, by reusing space, there exists a logspace algorithm that, given any set
$V$ of $N'$ vectors together with a ``list of choices'' $c\in\{0,1\}^l$ (with
$l=O(\log N')$), enumerates $l+1$ vectors $u^{(1)},\dots,u^{(l+1)}$ satisfying
the following condition~$(\star)$:
\begin{itemize}
\item $u^{(1)}$ is orthogonal to at least $N'/2-\sqrt{N'}/2$ and at most
  $N'/2+\sqrt{N'}/2$ vectors of $V$.
\item Let $V_i\subseteq V$ be the subset of all the vectors $v\in V$
  satisfying $\forall j\leq i,v.u^{(j)} = c_j$. Then the vector $u^{(i+1)}$ is
  orthogonal to at least $|V_i|/2-\sqrt{|V_i|}/2$ and at most
  $|V_i|/2+\sqrt{|V_i|}/2$ vectors of $V_i$.
\end{itemize}
Note that $|V_i|$ is roughly divided by 2 at each step, so the number of
steps is $O(\log N')$. In particular, since $s'$ and $N'$ are simply
exponential, the following lemma is easily derived by combining what precedes
with Lemma~\ref{lem_full1}.
\begin{lemma}\label{lem_full2}
  There is an algorithm using work space polynomial in $n$ which, on input
  $(i,j,k,c)$ in binary, outputs the $j$-th bit of $u^{(i)}\in\{0,1\}^{N'}$,
  where the vectors $u^{(1)},\dots,u^{(l+1)}$ satisfy condition $(\star)$ for
  the input consisting of:
  \begin{itemize}
  \item the set $V$ of the $N'$ (full) satisfiable sign conditions compatible
    with $T^{(k)}$,
  \item together with the list of choices $c\in\{0,1\}^l$.
  \end{itemize}
\end{lemma}

\begin{lemma}
  There exists a \unif\vpspacezero\ family $(h_n)$ satifsying, for real $\bar x$
  and boolean $(i,k,c)$:
  $$h_n(\bar x,i,k,c) = \prod_{j\in u^{(i)}} f_j(\bar x),$$
  where $u^{(1)},\dots,u^{(l+1)}$ are defined as in Lemma~\ref{lem_full2} (in
  particular they depend on $T^{(k)}$).
\end{lemma}

\begin{proof}
  Lemma~\ref{lem_full2} asserts that deciding whether $j\in u^{(i)}$ is done
  in polynomial space. The use of Lemma~\ref{lem_sum_prod_pspace} then
  concludes the proof. \qed
\end{proof}

Therefore, by a \unif\vpspacezero\ test, one is able to know the sign of the
polynomial $h_n(\bar x,i,k,c) = \prod_{j\in u^{(i)}} f_j(\bar x)$. As
mentioned before, this gives us the inner product of $u^{(i)}$ and the (full)
sign condition of $\bar x$: this sign is $<0$ if and only if the inner product
is 1. By beginning with $c=0\cdots 0$ (step 1), and at step $i\geq 2$ letting
$c_{i-1}=1$ if and only if the preceding test was $<0$, the number of sign
conditions that have the same inner products as that of $\bar x$ is divided by
(roughly) two at each step. At the end, we therefore have a list of choices
$c$ that only the sign condition of $\bar x$ fulfills. This proves the
following lemma.

\begin{lemma}\label{lem_full3}
  There is a polynomial-time algorithm with \unif\vpspacezero\ tests which on
  input $\bar x$ outputs the list of choices $c$ (defined as above) which
  uniquely characterizes the sign condition of $\bar x$, provided we know the
  rank $k$ of the truncated sign condition $T^{(k)}$ of $\bar x$.
\end{lemma}

We are now able to recover the rank of the sign condition of $\bar x$ from
this information, as explained in the next section.

\subsection{Recovering the Rank of the Sign Condition}\label{sec_recovering}

\begin{lemma}\label{lem_number1}
  There is an algorithm using work space polynomial in $n$ which, on input
  $c\in\{0,1\}^l$ (a list of choices) and $k$, outputs the rank of a
  satisfiable sign condition compatible with $T^{(k)}$ that fulfills the list
  of choices $c$.
\end{lemma}

\begin{proof}
  In polynomial space we recompute all the vectors $u^{(i)}$ as in
  Lemma~\ref{lem_full2}, then we enumerate all the sign conditions thanks to
  Theorem~\ref{th_renegar} until we find one that fulfills the list of choices
  $c$.\qed
\end{proof}

The proof of Theorem~\ref{th_number2} follows easily from
Proposition~\ref{prop_trunc} and Lemmas~\ref{lem_full3}--\ref{lem_number1}.

\subsection{A Polynomial-time Algorithm for $\mathrm{PAR}_{\rr}$
  Problems}\label{sec_algo}

Remember that $A\in\parr^0$ and $(C_n)$ is a uniform family of polynomial-depth
algebraic circuits deciding $A$.

\begin{lemma}\label{lem_accept}
  There is a (boolean) algorithm using work space polynomial in $n$ which, on
  input $i$ (the rank of a satisfiable sign condition), decides whether the
  elements of the $i$-th satisfiable sign condition $S$ are accepted by the
  circuit $C_n$.
\end{lemma}

\begin{proof}
  We follow the circuit $C_n$ level by level. For test gates, we compute the
  polynomial $f$ to be tested. Then we enumerate the polynomials
  $f_1,\dots,f_s$ as in Proposition~\ref{prop_slice} for the circuit $C_n$ and
  we find the index $j$ of $f$ in this list. By consulting the $j$-th 
bit of the
  $i$-th satisfiable sign condition with respect to $f_1,\dots,f_s$ (which is
  done by the polynomial-space algorithm of Theorem~\ref{th_renegar}), we
  therefore know the result of the test and can go on like this until the
  output gate.\qed
\end{proof}

\begin{theorem}
  Let $A\in\parrzero$. There exists a polynomial-time algorithm with
  \unif\vpspacezero\ tests that decides $A$.
\end{theorem}

\begin{proof}
  $A$ is decided by a uniform family $(C_n)$ of polynomial depth algebraic
  circuits. On input $\bar x$, thanks to Theorem~\ref{th_number2} we first
  find the rank of the sign condition of $\bar x$ with respect to the
  polynomials $f_1,\dots,f_s$ of Proposition~\ref{prop_slice}. Then we
  conclude by Lemma~\ref{lem_accept}.\qed
\end{proof}
Theorem~\ref{th_transfer} follows immediately from this result.
One could obtain other versions of these two results 
by changing the uniformity conditions or the role of constants.

\end{document}